# Standard audiogram classification from loudness scaling data using unsupervised, supervised, and explainable machine learning techniques

Chen Xu[a]*, Lena Schell-Majoor[a] and Birger Kollmeier[a]

[a]*Medizinische Physik and Cluster of Excellence Hearing4all, Universität Oldenburg, D-26111 Oldenburg, Germany*

Contact: Chen Xu

chen.xu@uni-oldenburg.de

Department of Medical Physics and Acoustics, Faculty VI

University of Oldenburg, 26111, Oldenburg, Germany

Standard audiogram classification from loudness scaling data using unsupervised, supervised, and explainable machine learning techniques

**Objective:** To address the calibration and procedural challenges inherent in remote audiogram assessment for rehabilitative audiology, this study investigated whether calibration-independent adaptive categorical loudness scaling (ACALOS) data can be used to approximate individual audiograms by classifying listeners into standard Bisgaard audiogram types using machine learning (ML).

**Design:** Three classes of ML approaches—unsupervised, supervised, and explainable—were evaluated. Principal component analysis (PCA) was performed to extract the first two principal components, which jointly explained more than 50% of the variance. Seven supervised multi-class ML classifiers were trained and compared, alongside unsupervised and explainable methods.

**Study Sample:** A large auditory reference database (N = 847) containing ACALOS data was used for model development and evaluation.

**Results:** The factor map showed substantial overlap between listeners, indicating that cleanly separating participants into six Bisgaard classes based solely on their loudness patterns is challenging. Nevertheless, the ML models demonstrated reasonable classification performance. Among the supervised classifiers, logistic regression achieved the highest accuracy.

**Conclusions:** The findings demonstrate that ML models can predict standard Bisgaard audiogram types—within certain limits—from calibration-independent loudness perception data. This approach may support future hearing aid fitting in remote or resource-limited settings without requiring a traditional audiogram.

**Introduction**

Smartphones have been shown to be a feasible and practical platform for conducting auditory tests such as pure-tone audiometry, speech-in-noise tests, and adaptive categorical loudness scaling (ACALOS) (Xu et al., 2024a; 2024b; Saak et al., 2024). However, one major challenge for smartphone-based auditory assessments is the lack of device calibration. Unlike traditional clinical tests, where equipment is carefully calibrated, mobile devices used in at-home testing are typically uncalibrated, which can lead to inaccurate results. Previous studies (e.g., Almufarrij et al., 2023; Xu et al., 2024d) have suggested that certain supra-threshold listening tests are relatively robust to calibration issues, as their outcome measures often depend on level differences rather than absolute sound levels. Building on this insight, the primary goal of this study is to investigate whether standard audiogram classes (as defined by Bisgaard et al., 2010) can be estimated using various machine learning models trained on data from uncalibrated supra-threshold tests—specifically, the ACALOS procedure.

   Bisgaard et al. (2010) introduced ten standard audiograms—often referred to as Bisgaard profiles—to support hearing aid fitting. These profiles include seven representing flat or moderately sloping hearing losses (N-type profiles) and three representing steeply sloping losses (S-type profiles). Each profile is assigned a number, with higher numbers indicating greater degrees of hearing loss; for example, N1 represents a mild, flat-to-moderately sloping loss. Classifying individuals according to the appropriate standard audiogram enables a coarse but meaningful characterization of their hearing loss. This classification also facilitates an initial hearing aid fitting, as demonstrated by recent advances in profile-based fitting strategies (Dreschler et al., 2008; Sanchez-Lopez et al., 2021, 2022; Wu et al., 2022). As such, deriving this parameter is valuable for both clinical and research applications in audiology.

Although the audiogram remains a cornerstone of auditory assessment, accurately measuring hearing thresholds requires controlled conditions—specifically, low background noise and calibrated equipment. Consequently, reliable threshold measurements are often not feasible in remote or unsupervised environments. Supra-threshold measures, such as categorical loudness scaling (CLS), which have been applied in remote smartphone-based testing using uncalibrated devices (Xu et al., 2024b; 2025b), may offer a viable alternative—provided they are sufficiently robust to background noise and rely on level-difference measures that are largely independent of absolute level calibration. This study therefore investigates whether CLS (Heller, 1985; Kollmeier, 1997) can provide a reasonably accurate and robust estimate of an individual's audiogram, categorized according to the Bisgaard profiles. Since the data set used here was collected under calibrated conditions, we introduced random level offsets drawn from a Gaussian distribution to simulate the effects of uncalibrated testing environments.

Loudness is the subjective perception of sound intensity (Zwicker and Fastl, 2013; Heller, 1985, Oetting et al., 2014; 2016). It is typically assessed using the categorical loudness scaling (CLS) test, in which individuals grade presented sounds using categories such as 'very soft, 'medium', or ' loud'. The standard methodology for measuring loudness is outlined in ISO 16832 (2006), while the parametric outcomes, such as slope and uncomfortable loudness levels (UCL), are detailed in Oetting et al. (2014). As loudness is a key audiometric parameter in hearing research, it has been incorporated into the 'auditory profile' and subsequently validated in multi-center studies (Dreschler et al., 2008; Van Esch et al., 2013). Because the audiogram influences the average shape of loudness functions across frequency, it remains to be examined whether this statistical dependence can be exploited to estimate the standard audiogram from uncalibrated CLS data. In this study, we assess the feasibility of such estimation using different machine learning algorithms.

Broadly, machine learning approaches can be categorized into three types: unsupervised, supervised, and explainable (see Mellor et al., 2018 for further details). The key distinction between unsupervised and supervised ML lies in the use of labels—unsupervised ML operates on unlabeled data. Initially, vector quantization (i.e., one unsupervised ML model) was employed to cluster clinical data and extract standard audiograms (Bisgaard et al., 2010). Another widely-used unsupervised method is principal component analysis (PCA), which reduces the dimensionality of complex data into principal components (PCs). In recent hearing research, PCA has been applied to analyze loudness measurement data (Wu et al., 2022) and in studies such as Sanchez-Lopez et al. (2021) to derive auditory profiles.

Supervised machine learning (ML) is a framework for analyzing labeled data, with classification being one of its primary tasks. Various classification algorithms, often referred to as classifiers, such as support vector machines (SVM), random forests (RF), and logistic regression (LR), have been developed based on different underlying mechanisms (see Kelleher et al., 2020; Murphy, 2012; Witten & James, 2013 for foundational classification theory). In hearing research, Ellis and Souza (2021) applied three classifiers—RF, SVM, and K-nearest neighbor (KNN)—to categorize patients using the Wisconsin Age-Related Hearing Impairment Classification Scale (WARHICS). Additionally, Lenatti et al. (2022) employed classifiers, including gradient boosting (GB) and decision trees (DT), to detect hearing loss from speech-in-noise test data. In clinical research, Gathman et al. (2023) predicted pure-tone averages (PTA) using a decision tree-based algorithm, incorporating demographic data, clinical factors, and subjective hearing status. While previous studies have utilized supervised learning algorithms to predict PTA, classify normal vs. impaired hearing, or categorize patients according to WARHICS, to our knowledge, no studies to date have explored whether standard audiograms can be accurately classified using these methods.

Explainable machine learning (ML) has garnered significant attention, as interpretability is a key concern when applying ML models in clinical practice (Babu et al., 2023). Clinicians often hesitate to adopt black-box models due to their lack of transparency. To address this, SHapley Additive exPlanations (SHAP) was introduced to explain how individual features influence a model's predictions (Lundberg et al., 2017, 2020). In the auditory field, Lenatti et al. (2022) were the first to apply SHAP to examine how features from speech-in-noise tests, such as the speech recognition threshold (SRT), contribute to distinguishing between normal hearing and hearing-impaired listeners. Additionally, Lenatti et al. (2022) employed two other post-hoc explainability methods: partial dependence plots (PDP) and feature permutation importance. To date, no study has explored the relationship between loudness-based supra-threshold parameters and standard audiograms using these post-hoc explainable ML techniques.

Therefore, the aim of this study is to leverage multiple state-of-the-art machine learning techniques to predict the respective standard audiogram class from loudness scaling data, evaluate their performance, and systematically compare their efficency. The models are designed for use in scenarios where device calibration cannot be ensured, such as smartphone-based hearing assessments (Xu et al., 2024a; 2024b; 2024c). The following three research questions are addressed:

1. RQ1: To what extent is it feasible to use uncalibrated ACALOS data for estimating the corresponding Bisgaard class?

2. RQ2: How can the statistical dependence between ACALOS data and the audiogram be quantified, and can machine learning methods be employed to systematically demonstrate this dependence?
3. RQ3: What are the strengths and limitations of the three machine learning methods examined in this study?

**Materials and methods**

*Description of the data set*

The data set used in this study was provided by Hörzentrum Oldenburg gGmbH. It is a superset of the publicly available Oldenburg Hearing Health Repository (OHHR; Jafri et al., 2024) and comprises a comprehensive array of auditory assessments, including, among others, a clinical audiogram, speech-in-noise tests, and loudness scaling tests. A detailed description of the data set can be found in Gieseler et al. (2017) and Jafri et al. (2025). The data were collected for clinical diagnostic purposes, and all measurements were unaided. Notably, the data set is neither representative for the general German civilian population nor does it represent a national cohort of patients with ORL-related health problems. Although previous studies (e.g., Lenatti et al., 2022; Ellis & Souza, 2021) have demonstrated that age is a significant predictor of hearing loss, we excluded demographic data from our analysis, as the focus of this study was to predict the standard audiograms based solely on loudness measures.

For this study, we extracted a subset of the larger database, focusing on the audiogram and loudness scaling data, only. The audiogram data set included measurements from N = 1199 participants, conducted by lab staff trained as hearing aid acousticians using a clinical audiometer in a sound-treated booth (IEC 60645-1, 2002).

Hearing thresholds were measured at 11 frequencies (0.125 to 8 kHz) for both ears, using HDA200 audiometric headphones.

The loudness scaling data set consisted of N = 648 participants, who completed the adaptive categorical loudness scaling task (Brand & Hohmann, 2002). Narrowband stimuli, consisting of one-third-octave-band low-noise noise, were presented monaurally at center frequencies of 1.5 and 4 kHz for both ears. According to Oetting et al. (2014), six variables were derived: $L_{2.5}$, $L_{25}$, $L_{50}$, $m_{high}$, $m_{low}$, and $L_{cut}$. $L_{2.5}$, $L_{25}$, and $L_{50}$ represent sound levels at 2.5, 25, and 50 categorical units (CU) on the fitted loudness function, roughly corresponding to the hearing threshold level (HTL), medium loudness level (MLL), and uncomfortable loudness level (UCL), respectively. The variables $m_{high}$ and $m_{low}$ represent the slopes of the upper and lower segments of the loudness function, while $L_{cut}$ defines the transition point between the two slopes.

*Preprocessing the data*

To increase the effective sample size, the data set was reorganized such that each participant's left and right ear data were treated as separate entries, following the recommendations of Dubno et al. (2013) and Lenatti et al. (2022). This restructuring doubled the number of records in the audiogram data set from 1199 to 2398. After excluding records with missing data, 2139 ear-specific entries remained.

Each record was then categorized in one of the standard audiograms as proposed by Bisgaard et al. (2010). This categorization was based on minimal RMSE. There are ten distinct classes: Seven flat or moderately sloping classes (N1, N2, N3, N4, N5, N6, N7) and three steeply sloping classes (S1, S2, S3). These audiograms corresponded to various degrees of hearing loss, with N1 and S1 representing very mild hearing loss, N2 and S2 representing mild hearing loss, N3 and N4 along with S3 representing moderate hearing loss, N5 and N6 representing moderate/severe hearing loss, and N7 representing severe to profound hearing loss.

The same data preprocessing steps—data augmentation followed by data cleaning—were applied to the loudness data. After cleaning, N = 1272 records were retained, while N = 23 records were excluded due to missing values. The two datasets (audiogram and loudness) were then merged using the unique patient identifiers and ear side. Participants who did not complete both auditory assessments were removed, leaving N = 1231 records. Additionally, participants with pure-tone average (PTA, i.e., the average hearing thresholds over 0.5, 1, 2, and 4 kHz) values below 20 dB HL (N = 316), typically classified as normal hearing, were excluded (Please note that participants with normal hearing were excluded, as they could not be assigned to any Bisgaard profile).

Further refinement involved removing records from classes N1, N5, N6, and N7, as each comprised less than 5% of the data—that is, fewer than 35 entries. This resulted in a final data set of N = 847 records distributed across 6 classes: N2, N3, N4, S1, S2, and S3. Figure 1A illustrates the class distribution, showing that class N3 had the highest percentage (32.5%), while class S3 had the lowest (6.5%). Figure 1B presents the standard audiogram for each class.

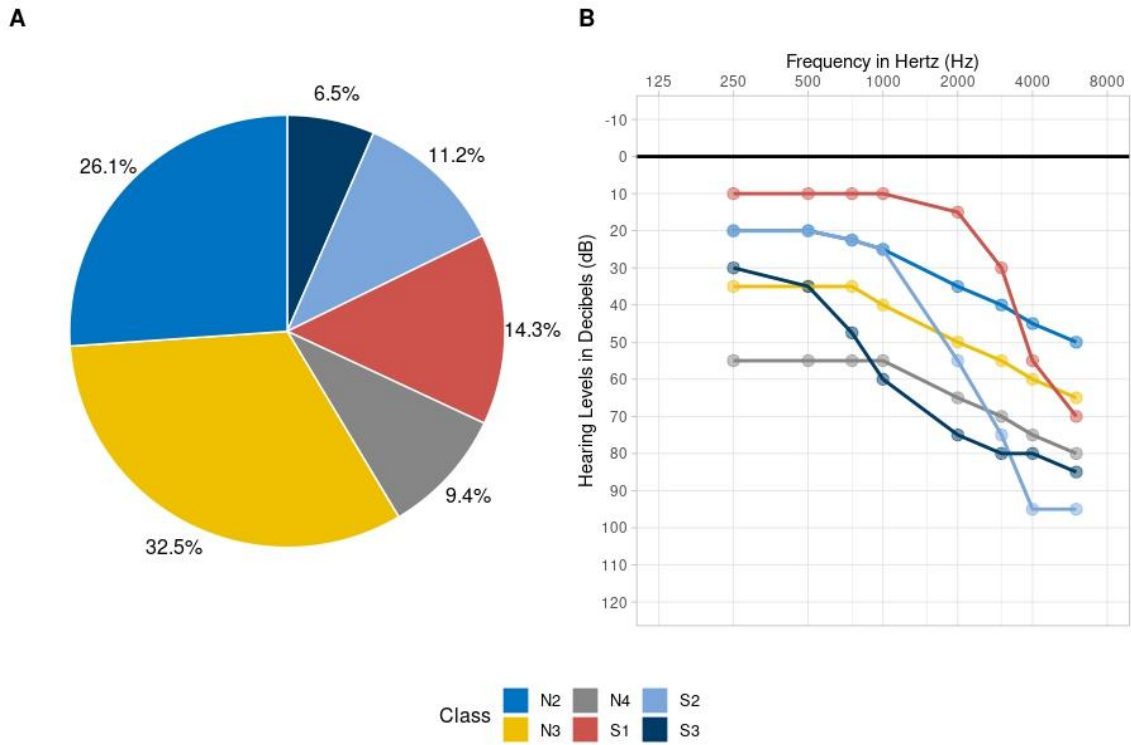

Fig. 1. A) pie chart showing the percentage of 6 classes in the final data set. B) original audiogram replicated from Bisgaard et al. (2010) (i.e., hearing thresholds in dB as a function of frequency) at frequencies ranging from 0.25 to 6 kHz of each class. N2, N3, and N4 are flat sloping groups whereas S1, S2, and S3 are steep sloping groups.

The descriptive statistics (i.e., first quartile, third quartile, mean, and standard deviation) for the processed features and labels from the loudness scaling data used in the machine learning models are summarized in Table S1. Twelve features were derived from the categorical loudness scaling test, while six labels were determined based on the clinical audiogram.

*Uncalibrated measurement simulation*

To simulate a data set with uncalibrated measurements, the same offset for both frequencies drawn from a Gaussian distribution with a mean of 10 dB (Scharf et al., 2024) and a standard deviation of 5 dB was applied to the features $L_{2.5}$, $L_{25}$, $L_{50}$, and

$L_{CUT}$ (please note that both ears of each participant had the same offset, as these parameters are likely influenced by device calibration). In contrast, the variables $m_{high}$ and $m_{low}$ were left unchanged, as they represent relative quantities and are thus independent of calibration.

*Machine learning models*

*Unsupervised machine learning*

Principal component analysis (PCA) was applied to reduce the dimensionality of the feature space, facilitating the exploration and interpretation of the high-dimensional data set. Two principal components (PCs), capturing the most significant information to explain the variability in the data, were extracted. We conducted PCA on the twelve features derived from the loudness scaling test using the R packages 'factoextra' and 'FactoMineR'. The former was employed for visualization, while the latter handled the analysis (Kassambara & Mundt, 2017; Lê et al., 2008). Prior to the PCA, the data set was standardized to unit variance, as per standard practice.

*Supervised machine learning*

Seven supervised machine learning algorithms—decision tree (DT), gradient boosting (GB), K-nearest neighbors (KNN), logistic regression (LR), neural network (NN), random forest (RF), and support vector machine (SVM)—were employed to classify the standard audiogram based on the twelve given features. Lenatti et al. (2022) provided detailed explanations for most of these algorithms, except for NN, which we additionally evaluated. All algorithms were implemented using the 'scikit-learn' package in Python (Pedregosa et al., 2011).

Before training, the input features were standardized so that the data distribution had a mean of 0 and a standard deviation of 1. To mitigate overfitting, a cross-validation procedure was applied: the data set was equally split into 10 folds, with models trained on 9 folds and tested on the remaining fold in each iteration. This process was repeated across 10 iterations, and the model's performance was averaged over all iterations.

Given that the classification task involved multiple classes, we used the One-vs-Rest (OvR) heuristic method to break the multi-class problem into multiple binary classification tasks (Murphy, 2012). For each class, a binary classification model was fitted, predicting whether the subject belonged to that class or not. The final class prediction was made by selecting the class with the highest probability (argmax). To implement the OvR approach, the multi-class labels were binarized by treating each class as the positive class in turn, while grouping all other classes as the negative class.

The DT classifier used 'entropy' as the criterion for measuring the quality of splits. The random state was set to 0 to ensure deterministic behavior during fitting, and a minimum of 5 samples was required to split an internal node. For the GB classifier, the hyperparameters—number of estimators, learning rate, and random state—were left at their default values (i.e., 100, 1.0, and 0, respectively), while the maximum depth of individual regression estimators was set to 2. In the KNN classifier, the number of neighbors was set to 2. The LR classifier also had a random state of 0.

We constructed a simple NN classifier with 1 input layer, 2 hidden layers (20 and 10 neurons, respectively), and 1 output layer. The 'lbfgs' solver was used for weight optimization, with a maximum of 3000 iterations. The random state was set to 1, and the alpha value controlling the strength of the $L_2$ regularization term was kept at its default value. In the RF algorithm, 10 trees were used to form the forest. Finally, C-supported vector classification was applied in the SVM classifier, with the regularization parameter set to 1000.

Two metrics, balanced accuracy and weighted $F_1$ score, were used to evaluate the overall performance of the seven classifiers mentioned earlier (Kelleher et al., 2020). Balanced accuracy (BA) is defined as:

$$BA = \frac{1}{2}(\frac{TP}{TP + FN} + \frac{TN}{TN + FP}) \tag{1}$$

where TP, TN, FP, FN denote True Positive, True Negative, False Positive, and False Negative, respectively. Balanced accuracy adjusts the standard accuracy score to account for imbalanced data sets. The $F_1$ score for each class is defined as:

$$F_1 = \frac{2TP}{2TP + FP + FN} \tag{2}$$

In the multi-class scenario, the weighted average $F_1$ score was computed. Both metrics range from 0 to 1, with higher scores indicating better classification performance.

Additionally, a confusion matrix, receiver operating characteristic (ROC) curve, and precision-recall curve were constructed to compare the local performance of a single classifier across different classes (Fawcett, 2006). In the confusion matrix, each row represents the true class, while each column denotes the predicted class. The value in the i-th row and j-th column indicates the number of instances with the true label of class i and the predicted label of class j. We normalized the values in the confusion matrix by the predicted (column) conditions.

To establish the ROC curve for each class, we plotted 1-specificity against sensitivity and computed the area under the curve (AUC). Specificity was calculated as TP / (TP + FN), while sensitivity was defined as TN / (FP + TN). The micro-average

OvR AUC score, which aggregates the contributions of all classes, was computed to reveal the average performance of the classifier. For the precision-recall curve, we first calculated precision and recall according to Schütze et al. (2008):

$$Precision = \frac{TP}{TP + FP} \tag{3}$$

$$Recall = \frac{TP}{TP + FN} \tag{4}$$

Later, the precision was plotted as a function of recall for each class to construct the precision recall curve, where the average precision (AP) score is calculated by:

$$AP = \sum_n (R_n - R_{n-1}) P_n \tag{5}$$

where $P_n$ and $R_n$ represent the precision and recall at the n-th threshold. The micro-average OvR AP score was calculated. Both the AUC and AP range from 0 to 1, with higher values indicating better classifier performance.

*Explainable machine learning*

Two model-agnostic explainable machine learning approaches—SHAP values and permutation feature importance—were employed for post-hoc analysis (Molnar, 2022). These explainability techniques aim to illustrate to what extent different features contribute to model outputs.

SHAP values (SHapley Additive exPlanations) are a game theory-based method for interpreting opaque machine learning models (Lundberg et al., 2017, 2020). In this study, a beeswarm plot was created as a summary visualization of the global effects of features using the 'shap' package in Python. In the beeswarm plot, each point represents a record, with its feature value indicated by color. The SHAP values for each feature are plotted along the x-axis. A large SHAP value for a point indicates a significant impact on the model output, while the top feature represents the most important feature.

The second approach applied was permutation feature importance, defined as the decrease in model accuracy when one feature is randomly shuffled (Fisher et al., 2019). A larger decrease in classification performance indicates greater importance of the feature. The algorithm was executed on both the training and test sets 10 times to permute each feature.

**Results**

*Unsupervised machine learning*

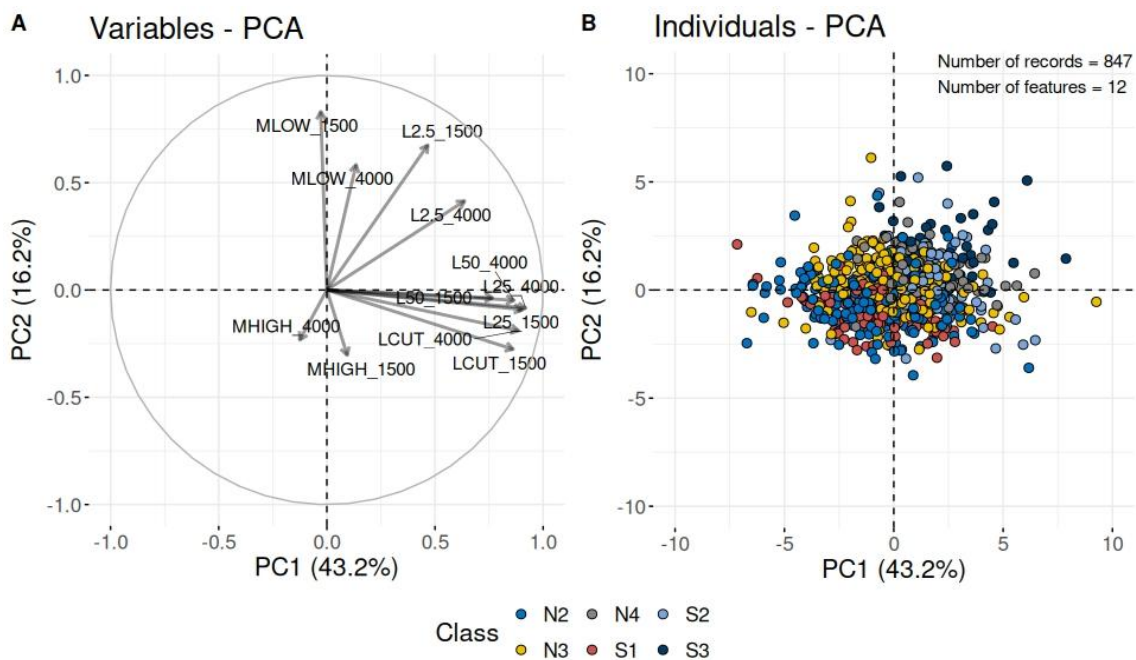

Fig. 2. Results of principal component analysis (PCA). A) PCA loading plot. Twelve loading vectors (arrows) denote twelve features. B) PCA score plot. Six classes are distinguished with different colors. x axis: score of 1$^{st}$ principal component (PC1). y axis: score of 2$^{nd}$ principal component (PC2). Please note that both panels A and B share the same axes (PC1 and PC2); however, in panel A, the vectors (loadings) are normalized and typically range between -1 and 1.

Figure 2 presents the PCA results for the processed loudness data. Panel A displays the loading vectors of individual features, while Panel B shows the sample scores for the first two principal components (PC1 and PC2), labeled according to six predefined classes (Please note that, due to an insufficient number of participants in the other four classes, only six classes were available for the PCA analysis). PC1 accounted for 43.2% of the total variance and PC2 for 16.2%, together explaining more than 50% of the total variance in the data set.

Most features positively loaded onto the PC1 while only $m_{low}$ at 1.5 kHz and $m_{high}$ at 4 kHz negatively loaded onto the PC1. Only $L_{2.5}$ and $m_{low}$ at two frequencies positively loaded onto PC2 while the remaining features negatively loaded onto PC2. The loading vectors at 1.5 kHz were not far away from those at 4 kHz though there were some differences in loading vectors between 1.5 kHz and 4 kHz. Furthermore, the loading vectors $m_{low}$ and $L_{2.5}$ were close to each other. These could be interpreted as the two features mostly contribute to the listening characteristics near threshold. Moreover, $L_{cut}$, $L_{25}$, and $L_{50}$ were neighboring as they seem to relate to the loudness perception at the medium-to-loud loudness level. Lastly, $m_{high}$ was observed outside the two main clusters, which likely indicates its limited variability across subjects and hearing loss types, and therefore its low predictive potential. The length of each loading vector indicates how much variance is accounted in the data set, i.e., features with longer vectors are more influential. Of all features, $m_{high}$ (closest to the origin) contributed the least to explaining data variability, whereas features such as $L_{25}$ and $L_{cut}$ were highly contributing to explained variance.

In the PCA score plot, the six subject classes showed significant overlap, with no clear boundaries to distinguish between them. Participants from the N4 and S3 classes predominantly scored positively on both PC1 and PC2, situating them in the first quadrant of the PC1–PC2 space (i.e., where both principal components have positive values). N3 subjects were dispersed across all quadrants, while S1 subjects generally had negative scores on PC2 (bottom half) and S2 subjects exhibited positive PC1 scores (right half). N2 subjects were distributed across the second, third, and fourth quadrants. Overall, individuals with moderate-to-severe hearing loss (e.g., N4 and S3) tended to show positive scores on both PC1 and PC2, whereas those with mild-to-moderate hearing loss (e.g., S1 and N2) displayed negative scores. This suggests that higher scores on both components are associated with increased hearing loss. Moreover, although PC1 and PC2 together explained approximately 60% of the variance, this may be insufficient to reliably distinguish among the six subject classes based solely on these two principal components.

*Supervised machine learning*

The performance of seven supervised machine learning classifiers was evaluated using balanced accuracy and the weighted $F_1$ score, as presented in Table 1. Most classifiers, with the exception of KNN and LR, achieved mean balanced accuracy and weighted $F_1$ scores above 0.9 on the training set. However, on the test set, mean values of both metrics ranged from 0.32 to 0.53 for all classifiers. As a result, all classifiers were considered at best moderately accurate in distinguishing the six classes. As anticipated, the training set exhibited higher balanced accuracy and weighted $F_1$ scores compared to the test set, with weighted $F_1$ scores being marginally higher than balanced accuracy scores. Overall, both metrics demonstrated consistent and stable performance across classifiers.

On the test set, LR achieved the highest mean balanced accuracy and weighted $F_1$ score, whereas KNN yielded the lowest values for both metrics. A pairwise t-test was conducted to compare the balanced accuracy and weighted $F_1$ scores among the classifiers on the test set. The results indicated that KNN significantly differed from GB, LR, NN, RF, and SVM in balanced accuracy ($p < 0.05$). For the weighted $F_1$ score, LR significantly differed from DT, KNN, and NN ($p < 0.05$). Finally, KNN significantly differed from GB and RF in the weighted $F_1$ score ($p < 0.05$). No significant differences were observed for the remaining comparisons.

Table 1. Performance of seven classifiers for the training and test data sets in terms of the balanced accuracy and weighted $F_1$ score. The largest and smallest balanced accuracy and weighted $F_1$ scores on the test set are highlighted. NN = Neural Network; GB = Gradient Boosting; LR = Logistics Regression; KNN = K Nearest Neighbor; SVM = Support Vector Machine; RF = Random Forest; DT = Decision Tree.

|  | balanced accuracy | | weighted $F_1$ score | |
| --- | --- | --- | --- | --- |
|  | Training | Test | Training | Test |
| DT | 0.90 (0.01) | 0.39 (0.06) | 0.92 (0.01) | 0.42 (0.05) |
| GB | 0.94 (0.16) | 0.44 (0.08) | 0.95 (0.14) | 0.48 (0.06) |
| KNN | 0.61 (0.01) | **0.32 (0.05)** | 0.69 (0.01) | **0.38 (0.05)** |
| LR | 0.51 (0.01) | **0.48 (0.07)** | 0.57 (0.01) | **0.53 (0.06)** |
| NN | 0.91 (0.02) | 0.40 (0.05) | 0.91 (0.02) | 0.42 (0.06) |
| RF | 0.98 (0.01) | 0.45 (0.05) | 0.98 (0.01) | 0.48 (0.05) |
| SVM | 0.95 (0.01) | 0.39 (0.05) | 0.95 (0.01) | 0.43 (0.04) |

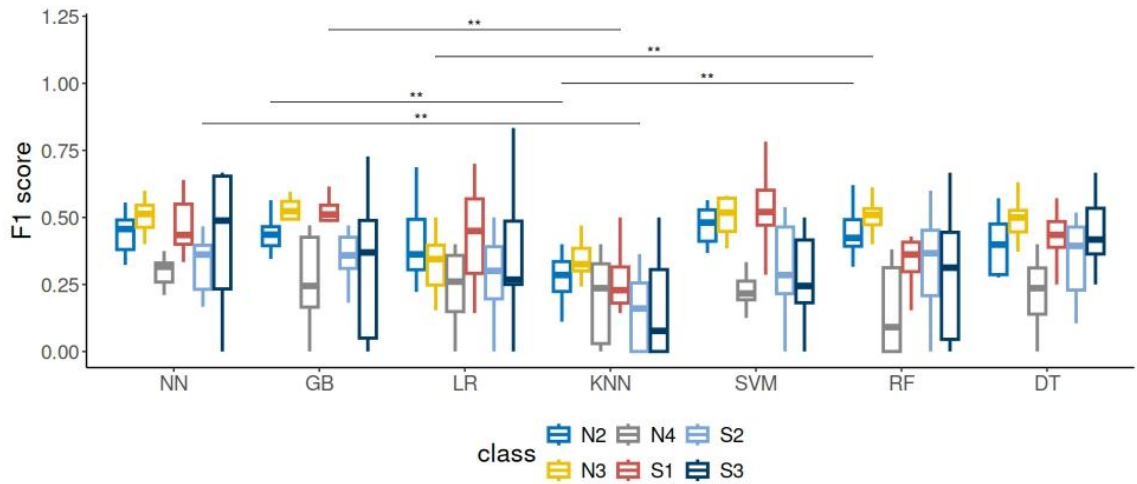

Fig. 3. $F_1$ score for each class as a function of the classifier on the test set. Refer to Table 1 for an explanation of the abbreviations for the classifier. The medians, 25%, 75% percentiles, and interquartile ranges (IQR) are given in the respective bar-and-whiskers plot. The ends of the whiskers describe values within 1.5*IQR of the 25% and 75% percentiles. The level of significance for p values, which indicates the difference between classifiers, is labeled with stars above the lines.

While Table 1 primarily summarizes the global performance of the classifiers (i.e., averaged across all 6 classes), Fig. 3 provides a more detailed view of the classification performance for each individual class, quantified by the $F_1$ score. The classifiers showed varying abilities to distinguish between the 6 classes. Notably, the N3 class had the highest median $F_1$ score across most classifiers, except for LR and SVM, indicating that N3 was generally classified with the highest accuracy. In contrast, LR and SVM performed best in classifying S1. The S3 class had the lowest median $F_1$ scores for KNN. Additionally, classifiers such as NN, GB, LR, SVM, RF, and DT struggled the most with predicting the N4 class. Overall, participants with mild-to-moderate hearing loss (e.g., N3, N2, and S1) were easier to classify, while those with more severe hearing loss (e.g., N4, S2, and S3) posed greater classification challenges. This suggests that greater hearing loss is associated with reduced variation in loudness scaling outcomes, making them less discriminative. In addition, the S3 class generally exhibited the largest interquartile ranges (IQRs) among all classes, particularly for the NN, GB, and RF models. In contrast, the N3 class showed relatively small IQRs.

     A pairwise t-test was conducted to compare the $F_1$ scores between classifiers. The results showed a significant difference for the N3 class between LR and RF ($p < 0.05$). Additionally, KNN significantly differed from RF and GB for the N2 class, NN for the S2 class, and GB for the S1 class ($p < 0.05$). No significant differences were found among the remaining classifier pairs.

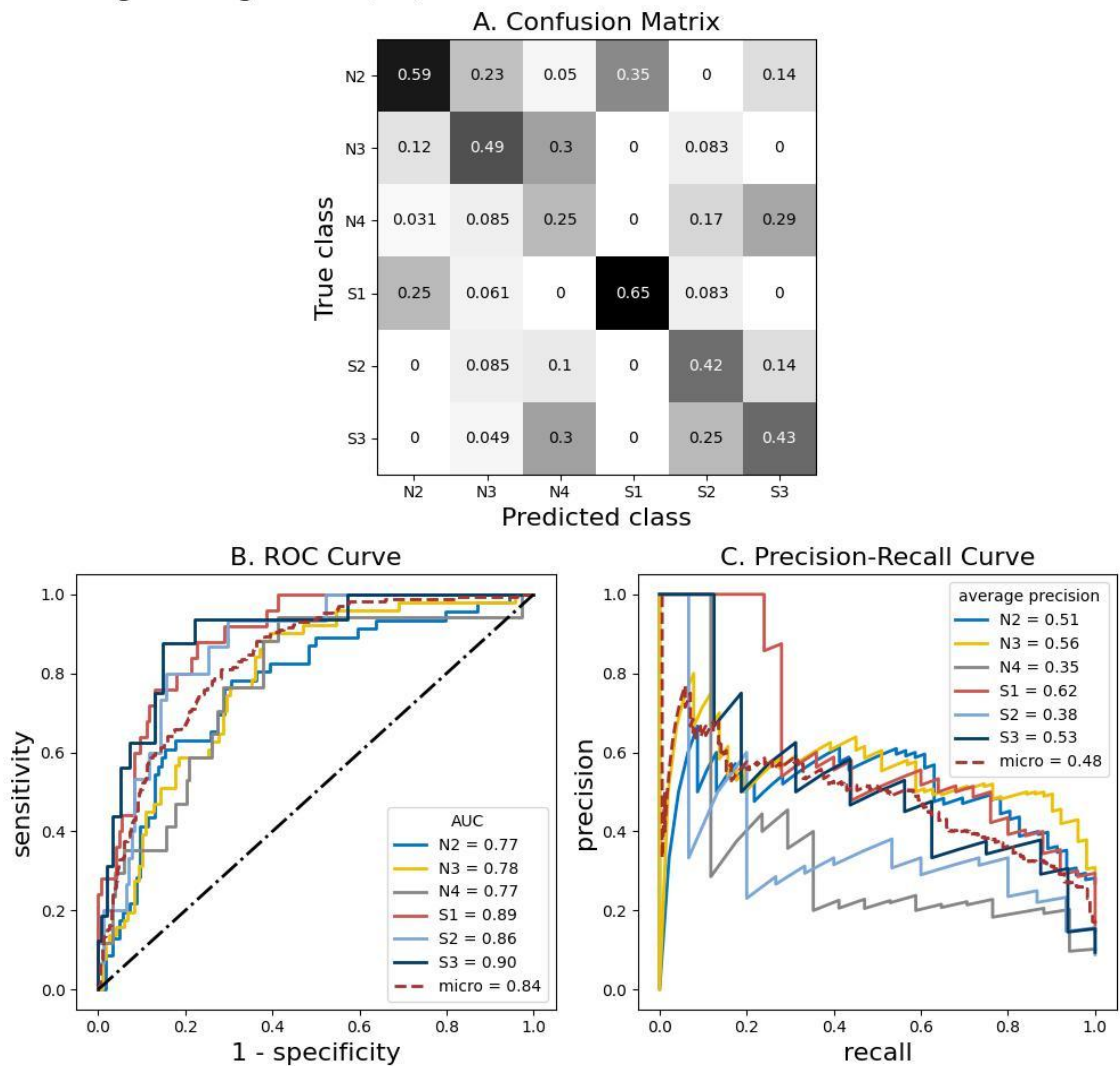

Fig. 4. A) Confusion matrix B) Receiver operating characteristic (ROC) curve. C) Precision recall curve for the **LR** classifier. AUC = Area under the curve; AP = average precision. Dashed lines: micro-average AUC and AP. Dash-dotted line in ROC curve: the baseline AUC (0.5).

Figure 4A presents the confusion matrix of the LR classifier across the six classes. Results from the LR classifier are presented, as it achieved the highest $F_1$ scores and balanced accuracy among all models. The intensity of the color corresponds to the magnitude of the values, which shows the normalized instances of combinations of the true and predicted class, with darker shades indicating larger values. As expected, the elements on the main diagonal represent the percentage of participants correctly classified. Values outside the main diagonal reflect misclassifications. As observed, the majority of the classes show the highest percentages along the main diagonal, indicating that the classifier performs well in predicting the standard audiograms. However, there is some confusion, particularly between the classes N4 and S3 and between N2 and S1.

Figure 4B shows the ROC curves for each class, with a baseline AUC of 0.5. Generally, the ROC curves for all classes, as well as the micro-average ROC, are above the baseline, indicating strong predictive performance for standard audiogram classification. Quantitatively, class S3 achieved the highest AUC value (0.90), suggesting near-perfect classification, while classes N2 and N4 had the lowest AUC (0.77), indicating relatively poor classification. Furthermore, the ROC curve for S3 is closest to the optimal point in the top-left corner, whereas N2 is farthest from it. The micro-average AUC was 0.84, further confirming that the overall classification performance across the six classes was satisfactory.

Figure 4C illustrates the precision-recall (PR) curves for each class. The PR curves display a zigzag pattern, with frequent fluctuations and intersections between classes. Notably, the precision-recall curve for S1 approaches the top-right corner, which represents optimal performance. Conversely, classes such as N4 and S2 are significantly distant from this point. The AP score for S1 was the highest (0.62), while N4 had the lowest AP score (0.35). The micro-average AP score was 0.48, suggesting that the LR classifier performed reasonably well in distinguishing between the classes.

Comparing Figures 4B and 4C jointly, it is evident that the LR classifier predicted S1 with relatively high accuracy, consistent with the high median $F_1$ score reported for this class in Figure 3, despite the large IQR of the $F_1$ score. Notably, the AP scores for all classes were lower than their corresponding AUC scores, which may be attributed to the imbalanced nature of the data set. According to Saito and Rehmsmeier (2015), AUC scores tend to remain unaffected by data imbalance, whereas AP scores decrease in imbalanced datasets. For instance, an AP score of 0.84 may indicate good early retrieval in a balanced data set, whereas in an imbalanced data set, an AP score of 0.51 is still considered reasonably good. Therefore, despite the AP scores being around 0.5 for both individual classes and the micro-average, the classifier's performance is still deemed strong, given the dataset's imbalanced nature.

*Explainable machine learning*

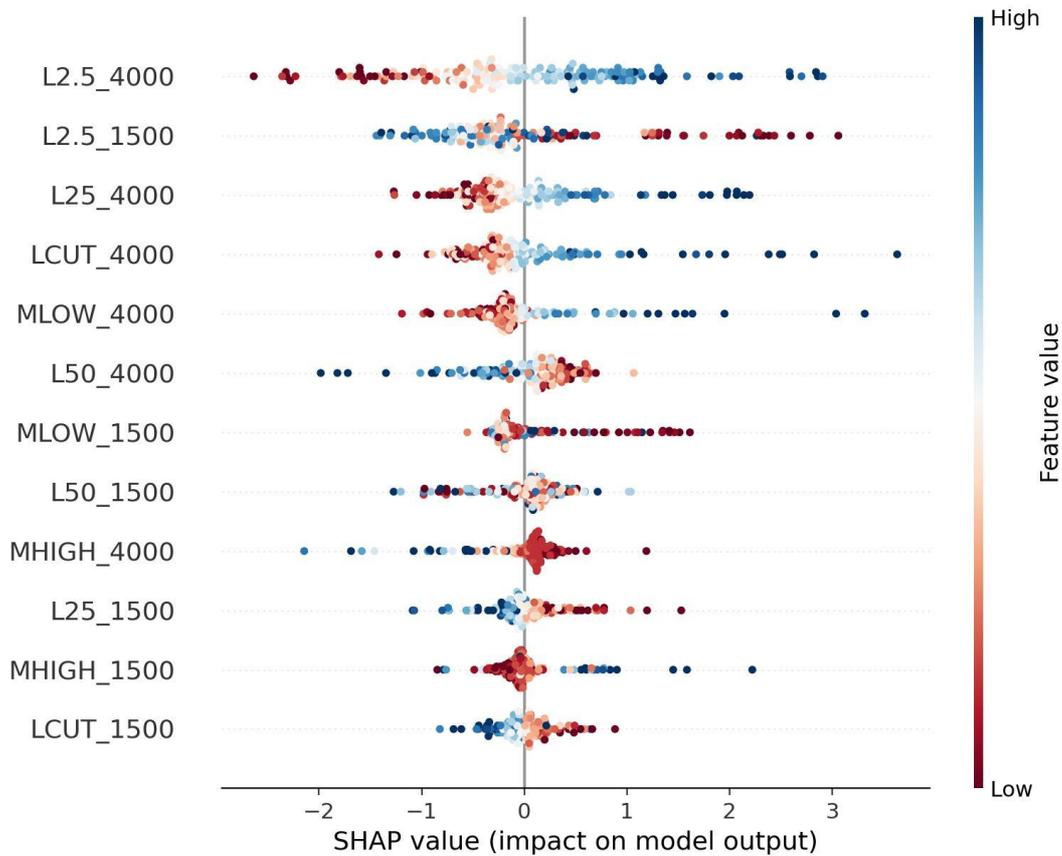

Fig. 5. Beeswarm plot. Features are listed from top to bottom based on their importance. X axis denotes the SHAP value, revealing the impact on model output. Each point represents one record, of which the color means the feature value.

Figure 5 displays a beeswarm plot, illustrating the impact of various features on the model's output. Features were ranked according to the sum of their SHAP values across all participants. The top three most important features were L2.5_4000, L2.5_1500, and L25_4000, whereas L25_1500, MHIGH_1500, and LCUT_1500 were the least influential. Starting with L2.5_4000, lower feature values were associated with more negative SHAP values, indicating a stronger contribution toward predicting the negative class. Note that negative SHAP values reflect the direction of influence, not poor model performance—both large positive and negative SHAP values indicate high feature importance. In contrast, higher feature values corresponded to more positive SHAP values, indicating a stronger contribution toward predicting the positive class. This can be interpreted as follows: individuals with low hearing thresholds at 4 kHz (represented by the red points in Figure 5) were less likely to be correctly classified into one of the six classes, while those with high hearing thresholds at 4 kHz (blue points) were more likely to be correctly classified into the standard audiograms. However, for the feature L50_4000, individuals with higher uncomfortable loudness levels at 4 kHz (blue points) were less likely to be correctly classified, whereas those with lower uncomfortable loudness levels at 4 kHz were more likely to be correctly classified. Lastly, LCUT_1500 had little influence on the model's output, regardless of its values, indicating that this feature is not useful for predicting the standard audiogram. Please note that in this study, we used class-agnostic SHAP values, which represent feature contributions averaged across all classes.

Overall, minimum loudness levels close to thresholds (i.e., $L_{2.5}$) at both 1.5 kHz and 4 kHz emerged as highly influential in predicting the standard audiogram. Moreover, features at 4 kHz, particularly $L_{2.5}$, $L_{25}$, $L_{50}$, and $m_{low}$, were generally more important than those at 1.5 kHz as they may play an important role in distinguishing N-type from S-type profiles. In contrast, the intersection levels between the low and high linear parts at 1.5 kHz (i.e., LCUT_1500) were found to be unimportant. Loudness discomfort levels (i.e., $L_{50}$) held medium importance in the model's predictions.

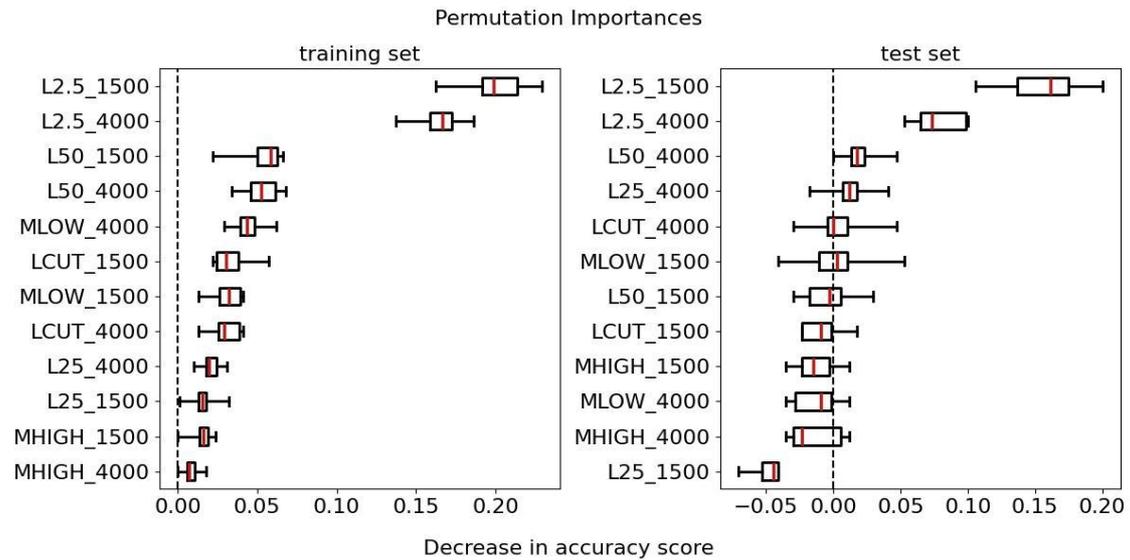

Fig. 6. Box plot of the permutation importance for the training set and the test set. The x axis denotes the decrease in accuracy score. The larger the value in x axis is, the more the feature leads to the decrease in accuracy score. The features are listed from top to bottom according to their importance. Dashed line: 0.

Figure 6 presents the results of the permutation importance analysis for both the training and test sets. L2.5_1500 and L2.5_4000 emerged as the two most influential features among the twelve features evaluated for both sets. In contrast, MHIGH_4000 was the least important feature for the training set, while L25_1500 was deemed the least important for the test set. The median importance scores for all features in the training set were greater than 0, indicating that each feature had a positive impact on predicting standard audiograms. However, for the test set, the median importance of L25_1500, MHIGH_4000, MLOW_4000, MHIGH_1500, LCUT_1500, and L50_1500 were below 0, suggesting that these features did not contribute meaningfully to classification in that set.

Interestingly, the results are not completely consistent with those using the SHAP approach (Figure 5), where L2.5_1500 was ranked more important than L2.5_4000. Additionally, these two features had a stronger influence on the training set compared to the test set. A disparity was also observed for other features; for instance, while L25_1500 was the least important feature in the test set, it ranked tenth in importance for the training set.

**Discussion**

This study applied three types of machine learning algorithms—unsupervised, supervised, and explainable—to predict standard audiograms from supra-threshold parameters derived from uncalibrated loudness scaling data. The main results are as follows: First, principal component analysis (PCA) revealed that although PC1 and PC2 explained ~60% of the variance, they were insufficient to reliably distinguish the six subject classes. Second, we evaluated the classification performance of supervised machine learning models for predicting audiogram profiles. Although the logistic regression (LR) model achieved the best performance among the tested models, its accuracy was limited (0.48 ± 0.07; see Table 1), and estimation errors remained non-negligible. Further details and supporting analyses are provided in the appendix. Third, we applied explainable machine learning methods to identify the most influential features and found that loudness-based threshold estimates at 1.5 kHz and 4 kHz emerged as the two most influential features for predicting the standard audiogram, while high-level slopes at both frequencies were of minor importance.

*Observations regarding the research questions*

With respect to the research questions (RQs) stated in the introduction, the following observations can be made:

**RQ1 (Feasibility).** Because of insufficient data in four Bisgaard classes, the analysis was restricted to six standard audiogram classes. Seven supervised machine learning classifiers trained on twelve loudness-scaling features showed moderate performance (balanced accuracy (BA) and weighted $F_1 \approx 0.45$). Logistic Regression achieved the highest scores ($F_1$ = 0.53, BA = 0.48), while K-Nearest Neighbors performed worst ($F_1$ = 0.38, BA = 0.32). These results indicate that audiogram prediction from uncalibrated CLS data is feasible, but with clear accuracy limitations.

**RQ2 (Statistical dependence between ACALOS data and the audiogram).** Principal Component Analysis (PCA) indicated that separating six classes within the space of the first two principal components is challenging, suggesting that unsupervised PCA alone is insufficient. In contrast, supervised ML revealed moderate covariation between loudness-scaling features and audiogram classes, particularly for mild-to-moderate hearing loss and calibration-independent features such as $m_{low}$. Explainable ML further highlighted L2.5_4000 and L2.5_1500 as the most influential features, with general agreement on highly and weakly important features but less consensus for features of intermediate importance. This high influence of loudness-based threshold estimates L2.5_4000 and L2.5_1500 on audiogram classification is not unexpected. Yet the uncalibrated measurement simulation with a random level offset did not completely remove the statistical dependence between the (indirect) threshold measures from loudness scaling and the audiogram (see below).

**RQ3 (Strengths and weaknesses of the ML methods).** PCA proved unsuitable for class separation, supervised classifiers achieved moderate but consistent performance, and explainable ML provided insights into feature importance and model interpretability. Together, these observations underline both the promise and the current limitations of ML-based classification of standard audiograms from loudness-scaling data.

*Statistical dependence between the audiogram and loudness scaling data*

A central issue in predicting audiograms from CLS data is the degree of statistical dependence between supra-threshold loudness growth and pure-tone thresholds. Prior studies have shown that specific CLS-derived features (particularly $L_{2.5}$) correlate strongly with pure-tone thresholds, whereas others (e.g., $m_{high}$) do not (Al-Salim et al., 2010; Oetting et al., 2014). Our results are consistent with this literature: L2.5_4000 and L2.5_1500 were most predictive, while $m_{high}$ features contributed negligibly. Please note that the data presented here include a random variable added to the absolute threshold values to simulate the influence of an unknown calibration offset. This variable was expected to increase variability in the thresholds and thereby reduce the correlation between the audiogram and loudness-scaling data. The results, however, indicate that this effect was only marginal (see Appendix for details), likely because the offset was assumed to be identical across all frequencies and ears for a given subject.

The moderate accuracy observed across classifiers (~0.45) reflects this limited dependence. Since loudness growth functions capture both threshold-related and supra-threshold dynamics, they cannot fully substitute pure-tone audiometry. Moreover, inter-individual variability in slopes and UCL values means that even perfect modeling of the CLS data will only approximate audiometric thresholds. These findings suggest the existence of a theoretical performance ceiling, where supra-threshold loudness measures constrain maximum achievable accuracy in predicting audiograms.

Nevertheless, rough audiogram classification still holds clinical value. Unlike digits-in-noise tests, which lack threshold information, CLS-derived features provide an approximation of audiogram shape, which can guide initial hearing aid fitting. Future work could improve performance by integrating additional features (e.g., demographic factors, self-reported hearing, or speech-in-noise data), as suggested by Lenatti et al. (2022) and Vercammen & Strelcyk (2025).

### *Comparison of ML techniques employed*

The three categories of machine learning applied here illustrate different perspectives on the problem.

1. Unsupervised learning (PCA) offered insights into the structure of the feature space, highlighting which features drove variability and revealing clusters of feature types. However, PCA projections were insufficient for classifying six groups, underscoring the complexity of the data and the limits of unsupervised clustering for this task.
2. Supervised classifiers directly addressed the prediction of audiograms. Logistic Regression emerged as the most effective model, consistent with literature showing that linear models often outperform complex ones in small-to-moderate datasets (<20k records; Kass, 2019). In contrast, KNN struggled with the curse of dimensionality, confirming that not all models are equally suited for this feature set.
3. Explainable ML methods bridged predictive modeling with audiological interpretation, identifying the most and least relevant features. This step is crucial for clinical translation, as it clarifies why models make certain predictions and builds trust among practitioners.

Overall, each ML paradigm contributes differently: unsupervised methods aid feature interpretation, supervised methods enable classification, and explainable methods provide transparency. These findings suggest that supervised methods (e.g., logistic regression) are preferable, as the overall limited performance of all algorithms leaves little justification for using explainable ML methods that typically yield even lower classification accuracy, despite their apparent advantages in interpretability.

***Limitations and outlook***

While remote testing using an adaptive categorical loudness procedure can yield valid estimates of supra-threshold hearing capabilities, its precision in approximating the audiogram from uncalibrated data appears to be very limited. As such, substituting remote audiogram assessment with this procedure as a general approach does not yet appear feasible. Nevertheless, the limited accuracy may still be sufficient for certain applications—for instance, providing a starting point for hearing-aid fitting and fine-tuning, where initial average settings derived from audiogram data are typically overridden by modern assisted or self-adjusted adaptive fitting procedures (e.g., Gösswein et al., 2023; Keidser et al., 2011). Moreover, loudness scaling data can be valuable for configuring specific dynamic compression parameters (e.g., Oetting et al., 2018). In addition, several modifications to the procedure may be warranted to enhance its accuracy, such as:

a) **Incorporating additional calibration measures** (e.g., Scharf et al., 2024) to address the remote calibration challenge. This would strengthen the link between calibrated audiometric thresholds and (at least approximately calibrated) supra-threshold loudness scaling data. Our data suggest that classification accuracy improves by 3% in the absence of calibration offsets (Please see the appendix). To evaluate the impact of calibration, we plan to collect data using uncalibrated equipment in future studies. This will allow us to validate the current approach—developed using calibrated equipment, where the absence of calibration was simulated by introducing random level offsets.

b) **Extending the frequency range** for loudness scaling by including additional test frequencies (e.g., 500 Hz, 6 or 8 kHz). This is expected to improve the accuracy of audiogram classification by capturing frequency-dependent variations in loudness perception more effectively and increasing robustness against random calibration errors.

c) **Adopting a modified version of the ACALOS procedure**, referred to as reinforced ACALOS (rACALOS; Xu et al., 2024c), which emphasizes measurements near the absolute threshold. This reinforcement aims to yield more accurate threshold estimates at the tested frequencies and, in turn, improve the accuracy of audiogram estimation via interpolation or extrapolation across frequencies.

While the present study employs cross-validation to split the training and test sets, testing the models on datasets provided by other institutions or clinics would offer a more reliable evaluation. An alternative approach for robust validation is to apply the machine learning models to (calibrated) open audiometric datasets, such as those provided by Sanchez-Lopez et al. (2021).

The basic principles of remote, uncalibrated testing for approximate audiogram classification could, in the future, be used to enhance large-scale cohort datasets—such as the National Health and Nutrition Examination Survey (NHANES), which reflects the listening characteristics of a broader population. Such methods may enable the estimation of individual hearing profiles without the need for clinical audiometry in every case, thereby providing an approximately calibrated representation of hearing ability at scale.

Recent studies (e.g., Gösswein et al., 2023) propose adaptive supervised, semi-supervised, or self-supervised strategies for hearing-aid fitting. These approaches require only a rough audiogram estimate to initialize the fitting, as the exact audiogram is not critical. The mapping between an audiogram and a standard fitting formula (e.g., NAL-NL2) is itself an approximation to the average case and must be adjusted to individual needs. Consequently, errors in the estimated audiogram affect only the initial settings, which are subsequently refined through iterative adaptation. After a few iterations, the influence of the initial fit largely disappears, making some inaccuracy in the initial audiogram acceptable if it reduces testing time. Hence, this suggests that our approaches may still be useful as an initial audiogram estimation step that could be incorporated into adaptive, remote hearing-aid fitting paradigms such as suggested by Gösswein et al. (2023).

**Conclusions**

From our study on the feasibility of using state-of-the-art machine learning techniques to predict standard audiograms from uncalibrated loudness scaling data, we can draw the following conclusions:

- In principle, it appears feasible – within clear limits - to determine an individual's audiogram class (e.g., the Bisgaard profile) without directly measuring hearing thresholds, but instead by using uncalibrated categorical loudness scaling data.

- Classification accuracy was generally limited, with the logistic regression (LR) model outperforming the six alternative machine learning methods.

- The most important features for classification were the levels corresponding to very soft loudness perceptions ($L_{2.5}$ at both frequencies), which are known from the literature to be imprecise estimators of hearing threshold when considered in isolation. These were followed by the uncomfortable loudness level at 4 kHz (L50_4000), which has been reported to co-vary with audiometric hearing loss.

- Increasing the level roving does not decrease the reliance on absolute-level measures (e.g., L2.5_4000), nor does it increase the relevance of relative measures (e.g., MLOW_1500). One possible reason is that the difference of the level-dependent features across frequency has been used for classification which would lead to a complete independence from level roving.

- Future work should experimentally assess uncalibrated loudness data, rather than relying on simulated random calibration offsets as done in the present study. Additionally, a reinforced ACALOS procedure (see Xu et al., 2024c) —with greater emphasis on accurately estimating threshold levels—appears to be a promising approach for improving the prediction accuracy of standard audiograms.

Taken together, this study demonstrates for the first time the feasibility of estimating audiogram classes from (simulated) uncalibrated supra-threshold categorical loudness data. The machine learning–based framework employed may be applicable to future mobile hearing assessments and hearing device fitting, particularly in scenarios where device calibration cannot be ensured and the requirements allow for a rather limited precision in audiogram estimation.

**Appendix (impact of different roving conditions on the standard audiogram classification)**

As shown in Figs A1 and A2, we conducted additional level roving conditions (i.e., calibration offsets) to complement the single fixed roving set presented in the main paper (see the test set in Fig. 4B and Fig. A1, and Fig. 6 and Fig. A2C), and compared these results to a reference condition without any calibration offset.

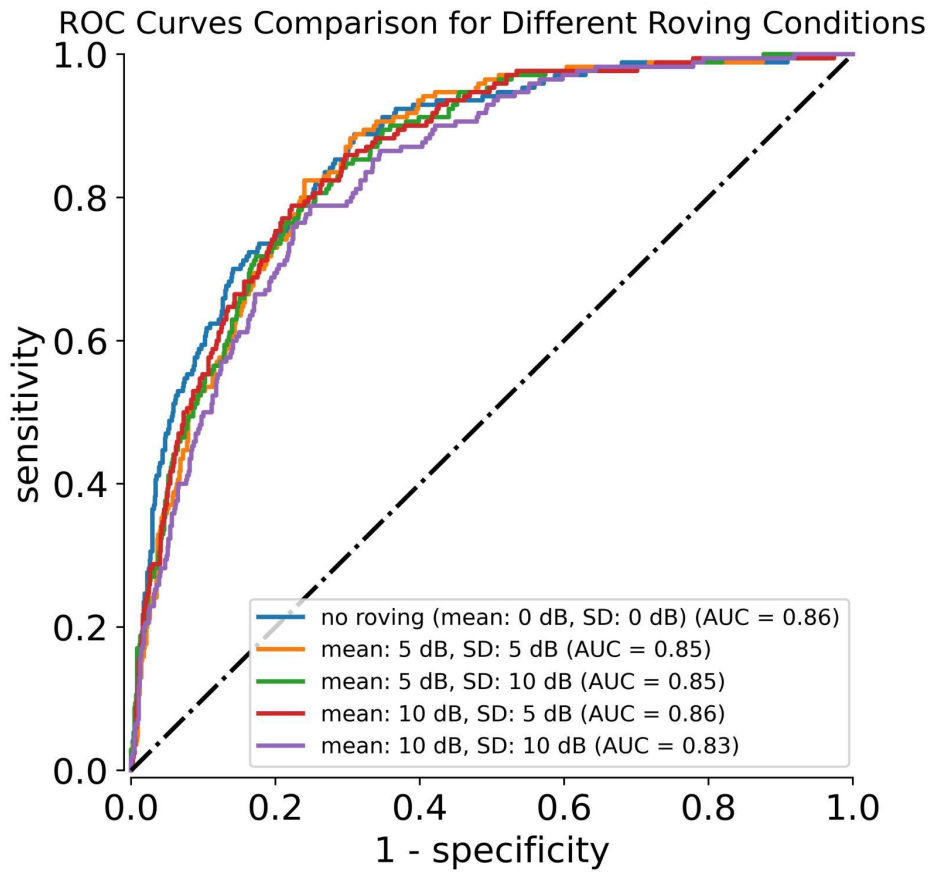

Fig. A1. ROC curves for five roving (calibration-offset) conditions: mean = 0 dB, SD = 0 dB (no roving as a reference condition); mean = 5 dB, SD = 5 dB; mean = 5 dB, SD = 10 dB; mean = 10 dB, SD = 5 dB; and mean = 10 dB, SD = 10 dB. See Fig. 4B for details on how the ROC curves are constructed. AUC (area under the curve) quantifies overall classification performance, with higher values indicating better discriminability.

Fig. A1 shows the ROC curves for the different roving conditions. Compared to the no-roving condition, the AUC values decrease slightly when level roving is introduced, indicating a modest reduction in accuracy. This trend is expected. However, the magnitude of the decrease is small across all roving conditions, suggesting that our AI models remain robust even when the amount of level roving increases.

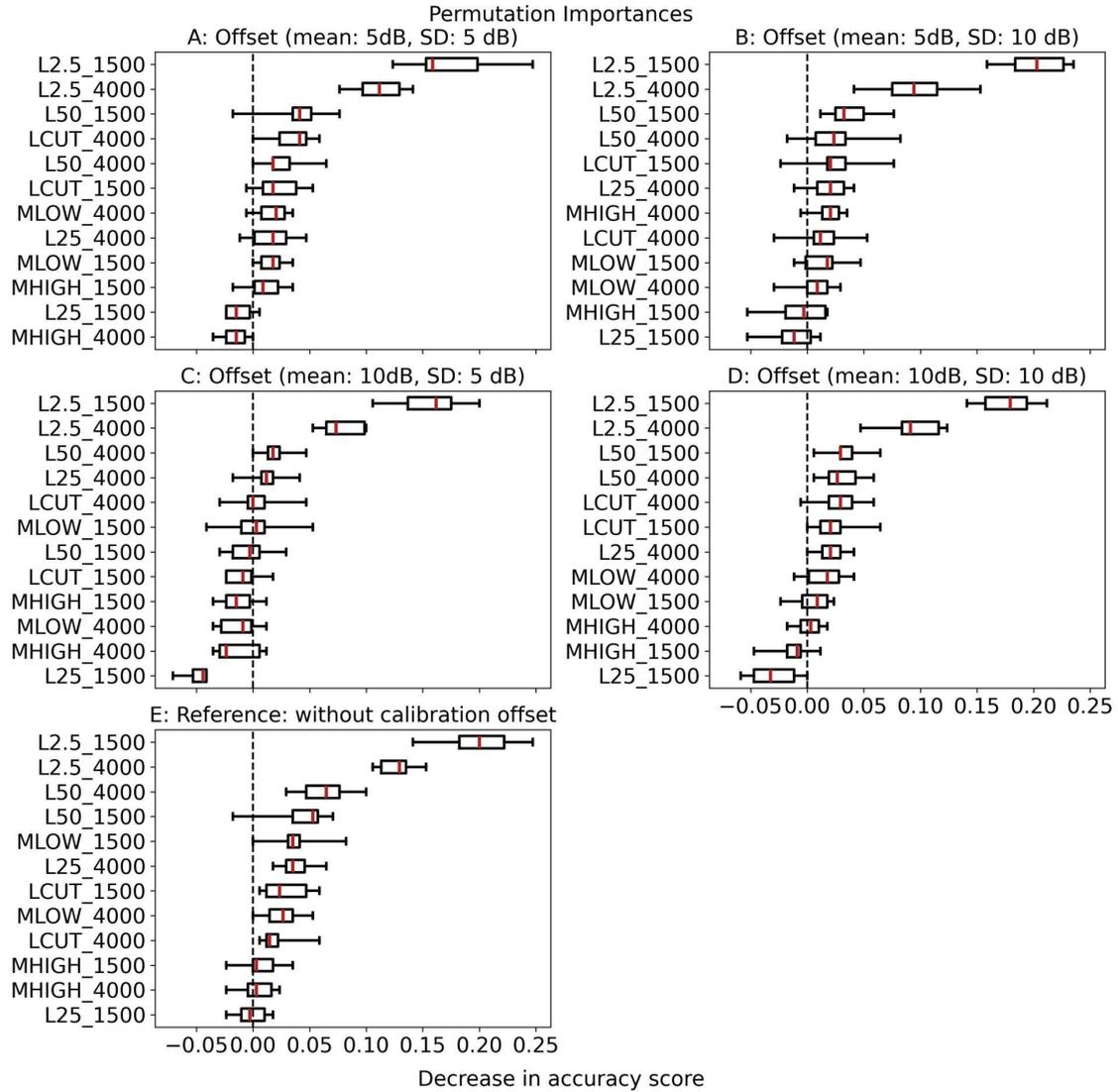

Fig. A2. Extension of Fig. 6 to compare feature importance across 12 loudness features under different calibration offset conditions: (A) mean = 5 dB, SD = 5 dB; (B) mean = 5 dB, SD = 10 dB; (C) mean = 10 dB, SD = 5 dB; (D) mean = 10 dB, SD = 10 dB. Subfigure (E) shows results without any calibration offset. For an explanation of the box plot, refer to Fig. 6. Note that subfigure (C) in Fig. A1 is identical to the test set results shown in Fig. 6.

Across all roving conditions shown in Fig. A2, absolute level features—particularly L2.5_1500 and L2.5_4000—consistently remained among the most important predictors, regardless of the magnitude of the roving applied.

In theory, increasing the level roving should reduce the importance of absolute level measures and increase the relevance of relative measures (also referred to as m-measures, such as $m_{low}$ and $m_{high}$). However, this expected shift was not as considerable in the experimental data as expected. Several possible explanations may account for this observation. First, the prediction targets—standard audiograms—may be correlated with level-dependent features (e.g., L2.5_1500), which could explain why these features remained informative for the model. One possible reason is that the machine-learning-based classification algorithm used the difference of the level-dependent features across frequency which would lead to a complete independence from level roving. Although relative measures were expected to play a more prominent role under level roving, the observed changes in classification accuracy were small and systematic rather than substantial.

Additional contributing factors may include:

(1) The roving range was too small to show a strong effect even though it was adapted in level range to the calibration offsets to be expected in real life.

(2) The machine learning algorithm is robust because it uses relative measures (m-values), reducing sensitivity to roving.

(3) The roving implementation (same roving pattern across ears/frequencies) limited the variability.

These possibilities will be explored in more detail in future work.

**Acknowledgments**

Parts of this study were presented at the 2024 annual meeting of the German Acoustical Society in Hannover (Xu et al., 2024d). This work was funded by the Deutsche Forschungsgemeinschaft (DFG, German Research Foundation) under Germany's Excellence Strategy – EXC 2177/1 - Project ID 390895286.

**Disclosure statement**

No potential conflict of interest was reported by the author(s).

**Glossary**

| Abbreviation | Meaning |
| --- | --- |
| AP | average precision |
| AUC | area under the curve |
| DT | decision tree |
| FN | false negative |
| FP | false positive |
| GB | gradient boosting |
| KNN | k nearest neighbour |
| LR | logistics regression |
| ML | machine learning |
| NN | neural network |
| OvR | One-vs-Rest |
| PC | principal component |
| PCA | principal component analysis |
| RF | random forest |
| SVM | support vector machine |
| TN | true negative |
| TP | true positive |